\documentclass[10pt,aps.prl,twocolumn,superscriptaddress,floatfix]{revtex4-1} 

\usepackage{amsmath}
\usepackage{graphics}
\usepackage{color}
\usepackage{caption}
\usepackage{subcaption}
\usepackage{float}
\usepackage{array}
\usepackage[margin=1.0in]{geometry}
\usepackage{ulem}
\usepackage{amssymb}

\begin{document}
\title{Stability of optimal-wavefront-sample coupling under sample translation and rotation}
\author{Benjamin R. Anderson$^1$, Ray Gunawidjaja$^1$, and Hergen Eilers$^{*}$}
\affiliation{Applied Sciences Laboratory, Institute for Shock Physics, Washington State University,
Spokane, WA 99210-1495}
\date{\today}

\email{eilers@wsu.edu}

\begin{abstract}
The method of wavefront shaping to control optical properties of opaque media is a promising technique for authentication applications. 
One of the main challenges of this technique is the sensitivity of the wavefront-sample coupling to translation and/or rotation. To better understand how translation and rotation affect the wavefront-sample coupling we perform experiments in which we first optimize reflection from an opaque surface -- to obtain an optimal wavefront -- and then translate or rotate the surface and measure the new reflected intensity pattern.  Using the correlation between the optimized and translated/rotated patterns we determine how sensitive the wavefront-sample coupling is.  These experiments are performed for different spatial light modulator bin sizes, beam spot sizes, and nanoparticle concentrations.  We find that all three parameters affect the different positional changes, implying that an optimization scheme can be used to maximize the wavefront-sample coupling's stability.  We also develop a model to simulate sample translation/rotation and its effect on the coupling stability, with the simulation results being 
qualitatively consistent with experiment.

\vspace{1em}

\end{abstract}

\maketitle

\vspace{1em}

\section{Introduction}
In 1990 Freund predicted that precise optical devices could be made using opaque media and wavefront shaping \cite{Freund90.01}.  Since then this technique has been used to control: transmission through opaque materials \cite{Vellekoop07.01}, the polarization of light \cite{Park12.01,Guan12.01}, broadband spectral characteristics \cite{Small12.01,Park12.02,Paudel13.01,Beijnum11.01}, and the spatio-spectral properties of random lasers \cite{Leonetti13.02,Leonetti12.02,Bachelard14.01,Bachelard12.01}. It has also been used to enhance fluorescence microscopy \cite{Vellekoop10.01,Wang12.01}, achieve perfect focusing \cite{Vellekoop10.01, Putten11.01}, compress ultrashort pulses \cite{McCabe11.01, Katz11.01} and enhance astronomical and biological imaging \cite{Mosk12.01,Stockbridge12.01}. Recently the technique's biological applications have been supplemented by photo acoustic wavefront shaping (PAWS) which combines wavefront shaping and ultrasound technology to better image biological tissue \cite{Lai14.02,
Chaigne14.01,Tay14.02,Gigan14.01,Tay14.01,Lai14.01}.

Additionally, using optimal wavefront shaping to control optical properties of opaque media has been proposed as a method of implementing optical physically unclonable functions (PUFs) \cite{Anderson14.05,Anderson14.06,Eilers14.01}.  PUFs are systems which use randomly distributed properties of a material (scatterers in opaque media, defects in semiconductor chips, etc) to create unique unclonable signatures \cite{Goorden13.01,Pappu02.01}.  The unique signatures of PUFs are of particular interest in authentication applications (such as authenticity verification and tamper indication) as their uniqueness makes counterfeiting a PUF unfeasibly difficult, if not impossible.

The envisioned application of optimal wavefront shaping to the implementation of optical PUFs is as follows: an opaque surface marker -- either intrinsic to the surface (such as defects or roughness) or an extrinsic glue (with scattering particles) -- is interrogated using a shaped wavefront and the surface's optical response is measured.  The response is then optimized using an optimization algorithm in order to determine an optimal wavefront. The optimal wavefront and surface response are then stored as a challenge-response pair, which is used at later times to verify that the surface is unchanged.  Changes in the challenge-response pair would indicate that the surface had been tampered with.

In an ideal implementation of this type of optical PUF, the changes in the challenge response pair would occur only due to intentional tampering.  However, in reality there are other unintentional effects which can change the challenge-response pair such as: dust accumulation, material changes due to the environment (temperature, humidity, radiation), drifts in the optical equipment, and sample positioning changes.  These various effects can work together to decouple the challenge and response, leading to a false indication of intentional tampering. Given these concerns it is necessary to quantify these unintentional effects if this technique is to be used in the field. 

In this study we specifically quantify the effects on the challenge-response coupling due to changes in sample positioning, including translation and rotation.  For our PUF samples we use a nanoparticle (NP) doped polymer and for the optical response we measure the reflected intensity pattern, with optimization working to focus the scattered light into a target area. In addition to experimentally measuring the effects due changes in sample positioning, we also use an extended random phase Gaussian beam model \cite{Anderson14.06} to simulate translation and rotation in order to better understand the underlying physical mechanisms.

\section{Theory}

\subsection{Extending the Random Phase Gaussian Beam Model}
Previously, we developed a random phase Gaussian beam model (RPGBM) \cite{Anderson14.06} to model optimized transmission through an opaque material.  One of the weaknesses of this model was the treatment of the material as only affecting the phase of the incident light, and not the amplitude. In order to take these effects into account we use a random matrix theory (RMT) model of scattering in disordered media \cite{Wigner51.01,Goodman00.01,Garcia89.01,Beenakker97.01,Pendry90.01,Pendry92.01,Mello88.01,Mello88.02,Molen07.01,Vellekoop07.01,Vellekoop08.01}. In the RMT model, scattering in a disordered media is represented by complex-valued matrices which map the incident electric field at input channel $m$ to the exiting electric field at output channel $n$. Specifically, using the RMT formulation of Vellekoop and Mosk, we can write the electric field at the output channel $n$ as

\begin{align}
E'_{t;n}=\sum_{m=0}^{N}t_{nm}E_m, \label{eqn:trans}
\\ E'_{r;n}=\sum_{m=0}^{N}r_{nm}E_m,
\label{eqn:ref}
\end{align}
with $E_m$ being the incident electric field at input channel $m$, $t_{nm}$ are the transmission matrix elements, which maps the $m^{th}$ input channel to the $n^{th}$ transmitted channel, and $r_{nm}$ are the reflection matrix elements, which map the $m^{th}$ input channel to the $n^{th}$ reflected channel. In this study we define the input and output channels to be related to the spatial coordinate on the sample surface, with the coordinates discretized for calculation purposes.

In Vellekoop and Mosk's model they only considered the transmission matrix elements \cite{Vellekoop08.01}, while in this study we are concerned with reflection.  Therefore at this point we will derive the reflection matrix elements. Assuming no absorption, the transmission matrix, $\mathbf{t}$, and reflection matrix, $\mathbf{r}$ are dependent on each other via

\begin{align}
\mathbf{r}^\dagger\mathbf{r}+\mathbf{t}^\dagger\mathbf{t}=\mathbf{I},
\label{eqn:conserv}
\end{align}
where $\mathbf{I}$ is the identity matrix and $\dagger$ denotes the conjugate transpose. The form of Equation \ref{eqn:conserv} arises due to conservation of energy in the absence of absorption.  From Vellekoop and Mosk's model we know that the transmission matrix can be modeled as a random unitary matrix, $\mathbf{U}$, drawn from a circular Gaussian distribution and scaled by a scalar transmission coefficient, $t$, giving the transmission matrix \cite{Vellekoop08.01,Beenakker97.01}:

\begin{align}
\mathbf{t}=t\mathbf{U},
\label{eqn:tmat}
\end{align}
where the form of Equation \ref{eqn:tmat} is a consequence of treating the scattering system using RMT.

Substituting Equation \ref{eqn:tmat} into Equation \ref{eqn:conserv} and rearranging terms we can derive an equation for the reflection matrix

\begin{align}
 \mathbf{r}^\dagger\mathbf{r}=\mathbf{I}-|t|^2\mathbf{U}^\dagger\mathbf{U}, \\
 \mathbf{r}^\dagger\mathbf{r}=(1-|t|^2)\mathbf{I},
\end{align}
which can be solved with a reflection matrix defined as

\begin{align}
 \mathbf{r}\equiv r\mathbf{V},
\end{align}
where $\mathbf{V}$ is a unitary matrix and $r$ is a scalar reflection term satisfying,

\begin{equation}
 |r|^2=1-|t|^2.
\end{equation}
Since the reflection matrix terms should be independent for a disordered medium \cite{Vellekoop08.01,Goodman00.01,Mello88.02,Beenakker97.01}, for our modeling we can draw the random unitary matrix $\mathbf{V}$ from a circular Gaussian distribution. This choice is also consistent with the maximal fluctuation theorem, which predicts that all eigenchannels are either open or closed \cite{Pendry90.01,Pendry92.01,Vellekoopthesis}.

\subsection{Simulations}
To model the effect of translation and rotation on the coupling between an optimal wavefront and sample we begin by assuming a Gaussian beam incident on the sample with an electric field of:

\begin{align}
E_m=E_0\exp\left\{-\frac{(m-m_0)^2}{\sigma^2}-i\psi_m\right\},
\label{eqn:inc}
\end{align}

where $m_0$ is the peak position, $\sigma$ is the Gaussian width, and $\psi_m$ is the phase mask due to the SLM.  Substituting Equation \ref{eqn:inc} into Equation \ref{eqn:ref} and applying the RPGBM we find the electric field in the detector plane to be:
\begin{align}
E_{d;n'}=\sum_{n=0}^N\sum_{m=0}^{N}r_{nm}E_0\exp\bigg\{-\frac{(m-m_0)^2\Delta x^2}{\sigma^2} \nonumber
\\ -i\frac{2\pi}{N}nn'-i\psi_m \bigg\},
\label{eqn:diff}
\end{align}
where $\Delta x$ is the computational grid spacing, $N$ is the total number of grid spaces, and the term $i\frac{2\pi}{N}nn'$ is the exponential term in the Discrete Fourier Transform.  Using Equation \ref{eqn:diff} and a microgenetic algorithm \cite{Anderson15.02} we systematically vary $\psi_m$ and calculate the intensity in the detector plane

\begin{align}
I_{d;n'}=|E_{d;n'}|^2,\label{eqn:int}
\end{align}
to find the phase mask giving a focused spot in the detector plane. At this point we note that during experiments we typically bin the SLM pixels into super-pixels of side length $b$ (i.e. the bin area is $b \times b$ px$^2$). For our 1-D simulations we still use bins but with them corresponding to groupings of grid points.

Once the optimized wavefront is found we use numerical transformations, discussed below, to model sample translation and rotation.  For each positional change we recalculate the intensity profile in the detector plane and compare the new profile, $I_i$, to the original optimized profile, $I_{0,i}$ using the Pearson correlation coefficient:

\begin{align}
c \equiv \frac{\sum\limits_{i=0}^N(I_{i}-\overline{I})(I_{0,i}-\overline{I_0})}{\sqrt{\sum\limits_{i=0}^N(I_{i}-\overline{I})^2\sum\limits_{i=0}^N(I_{0,i}-\overline{I_0})^2}},
\label{eqn:corr}
\end{align}
where $I_{0,i}$ is the initial optimized profile, $I_i$ is the intensity profile after translation or rotation, $\overline{I_0}$ is the average value of $I_{0,i}$, and $\overline{I}$ is the average value of $I_i$.  The coefficient varies from -1 to 1 with 1 corresponding to perfect correlation and -1 corresponding to perfect negative correlation. When performing simulations we use one spatial dimension, due to computational restraints, and Equation \ref{eqn:corr} is used as written with only one index.  However, in our experimental results we use two dimensions, which requires Equation \ref{eqn:corr} to be extended with a second index and all sums are turned into double sums.

\subsection{Translational Stability}

\subsubsection{Transverse Translation}
\begin{figure}
 \centering
 \includegraphics{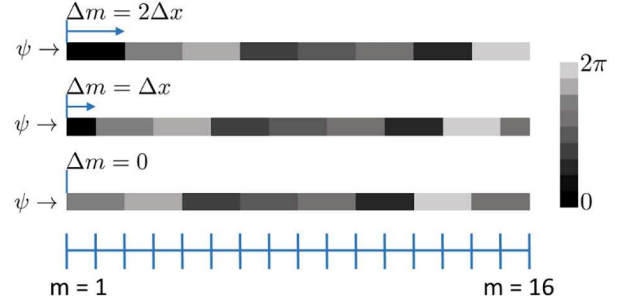}
 \caption{(Color online) Schematic diagram of transverse translation modeling using a grid with 16 points and a bin size of $b=2$. As the translation, $\Delta m$, is increased the wavefront shifts to the right, with the left wavefront bins becoming zero, and the right wavefront bins going off grid.  Note that while not shown the amplitude shifts as well.}
 \label{fig:XT}
\end{figure}

Once the optimal wavefront, $\psi_m$, has been determined and the optimal intensity profile recorded we can model transverse sample translation by shifting the incident field's index by $\Delta m$, changing the electric field in the detector plane to be:  

\begin{align}
E_{d;n'}=\sum_{n=0}^N\sum_{m=0}^{N}r_{nm}E_0\exp\bigg\{-\frac{(m+\Delta m-m_0)^2\Delta x^2}{\sigma^2} \nonumber
\\ -i\frac{2\pi}{N}nn'-i\psi_{m+\Delta m} \bigg\}.
\label{eqn:dx}
\end{align}
The new intensity profile is then calculated using Equation \ref{eqn:int} and is compared to the untranslated profile using Equation \ref{eqn:corr}. Figure \ref{fig:XT} shows a sketch of the wavefront for different translation values with the sample grid marked out and a bin size of 2$\Delta x$.

\begin{figure}
 \centering
 \includegraphics{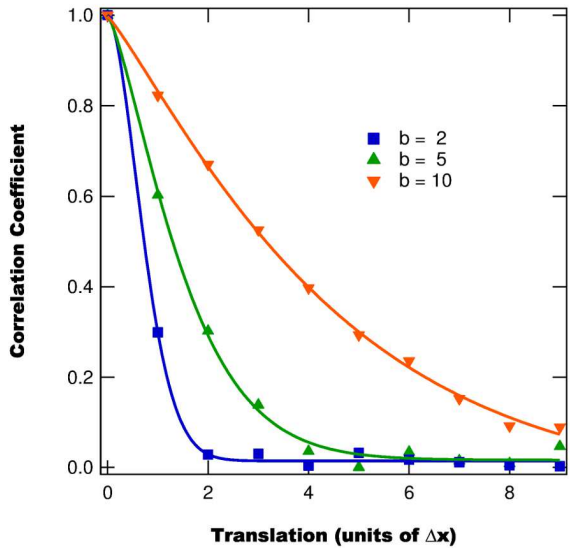}
 \caption{(Color online) Simulated correlation coefficient as a function of sample translation for a Gaussian width of 200 $\Delta$x and three different bin sizes with fits to a stretched exponential function.  As the bin size decreases the width of the stretched exponential decreases.}
 \label{fig:MX}
\end{figure}

\begin{figure}
 \centering
 \includegraphics{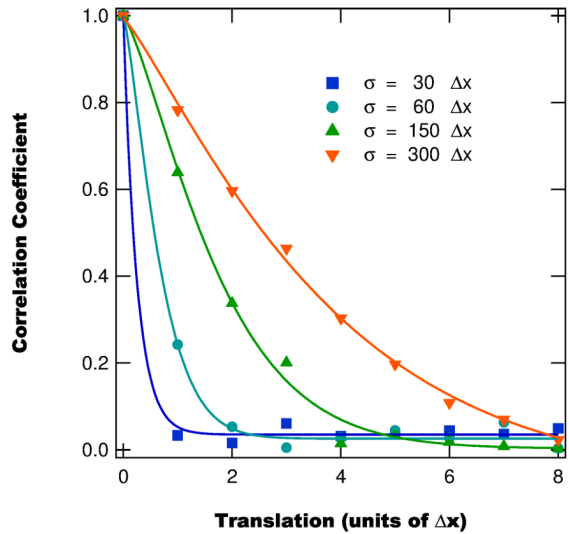}
 \caption{(Color online) Simulated correlation coefficient as a function of sample translation for a bin size of $b=5$ and four different beam spot sizes with stretched exponential fits.  As the spot size decreases the width of the stretched exponential decreases, suggesting that smaller spot sizes result in less coupling stability.}
 \label{fig:MSX}
\end{figure}

Using this method we calculate the correlation coefficient as a function of transverse translation for three different bin sizes, shown in Figure \ref{fig:MX}, and four different incident spot sizes, shown in Figure \ref{fig:MSX}.  The model results, as a function of translation, are found to follow a stretched exponential,

\begin{align}
 c(x)=A\exp\left\{-\left(\frac{x}{\sigma_X}\right)^\beta\right\}+c_0 \label{eqn:strexp}
\end{align}
where $c_0$ is a small offset from zero, $A=1-c_0$, and the functions width, $\sigma_X$, and exponent, $\beta$ depend on the bin and spot size.  Equation \ref{eqn:strexp} has a Half-Width-Half-Max (HWHM) of $W=\sigma_X(\ln 2)^{1/\beta}$.

From Figure \ref{fig:MX} we see that as the bin size increases the HWHM of the correlation curve increases implying greater resistance to decorrelation as the sample is translated.  This can be understood as the fractional translation is inversely proportional to the bin size (i.e. a translation of 2 $\Delta x$ is a complete bin shift for $b=2$, while it is only 1/5 of a bin shift for $b=10$), meaning the change due to translation is a smaller effect on the wavefront.

The other factor determining the fractional translation amount is the spot size of the incident beam, with the fractional translation being inversely proportional to the beam width. Figure \ref{fig:MSX} shows the correlation curves for different spot sizes, with the larger spot sizes resulting in greater HWHMs.  As with bin size, we find that the larger the segments of the phase front, the more resistant to transverse translation the wavefront coupling becomes. While these results suggest greater stability for larger bin sizes and spot sizes, the benefit needs to be weighed against the smaller enhancements that result when increasing the bin size or spot size \cite{Anderson14.06}.

\subsubsection{Translation Along Optical Axis}
\begin{figure}
 \centering
 \includegraphics{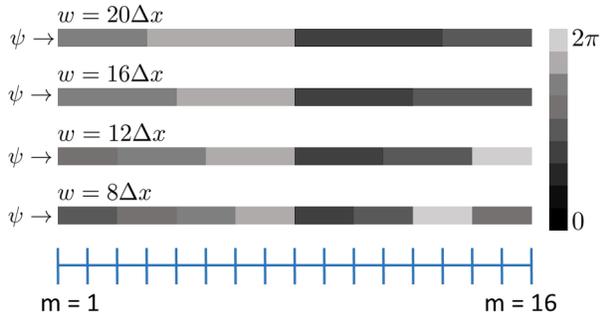}
 \caption{(Color online) Schematic diagram of modeling translation along the optical axis using a grid with 16 points. As the sample is translated along the optical axis the on sample beam width changes, which changes the coupling between the wavefront and sample grid.  Namely, as the beam width increases, the absolute bin size increases.}
 \label{fig:ZT}
\end{figure}

In addition to translating the sample in the transverse axis after optimization, we can also translate it along the optical axis resulting in the incident spot size changing. We model this effect by transforming the beam width, $w_0 \rightarrow gw_0$, and bin size, $b \rightarrow gb$ --such that the ratio $w/b$ is unchanged -- as well as introducing the effect of the changing Gaussian beam curvature given by

\begin{align}
R(g)=z_r\sqrt{g^2-1}\left(1+\frac{1}{g^2-1}\right)
\end{align}
where $z_R=\frac{1}{2}kw_0^2$ is the Rayleigh range and $k=2\pi/\lambda$, with $\lambda$ being the wavelength of the light used. Figure \ref{fig:ZT} shows a sketch of the wavefront for different spot sizes, with the bin size increasing with larger spot sizes.  Note that Figure \ref{fig:ZT} does not include the phase related to the Gaussian curvature, nor does it show the amplitude profile, which also changes with spot size.

Since we experimentally measure the change in $z$ position, not the change in spot size, we can convert from the spot size increase, $g$, to change in $z$ position by considering the Gaussian beam width,

\begin{align}
 w(z)=w_0\sqrt{1+\left(\frac{z}{z_R}\right)^2},
\end{align}
which for a new width, $gw_0$, can be rearranged to give the $z$ position:

\begin{align}
z&=z_R\sqrt{g^2-1}, \label{eqn:zpos}
\end{align}

\begin{figure}
 \centering
 \includegraphics{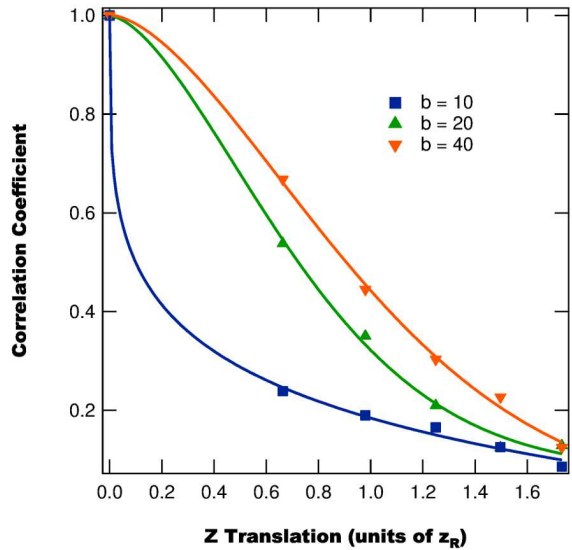}
 \caption{(Color online) Simulated correlation coefficient as a function of sample translation for a beam with a Gaussian width of 100 $\Delta$x and three different bin sizes with stretched exponential fits.  The correlation curves HWHM increases with large bin sizes.}
 \label{fig:MZ}
\end{figure}

Using the changes in the incident electric field described above we model the effect of translations along the optical axis on the wavefront coupling.  Figure \ref{fig:MZ} shows the correlation coefficient as a function of $z$ translation for three different bin sizes, with the HWHM increasing with larger bin sizes.  This result is similar to the transverse translation case, with the explanation also being that larger bin sizes reduce the fractional wavefront change due to translation.

\begin{figure}
 \centering
 \includegraphics{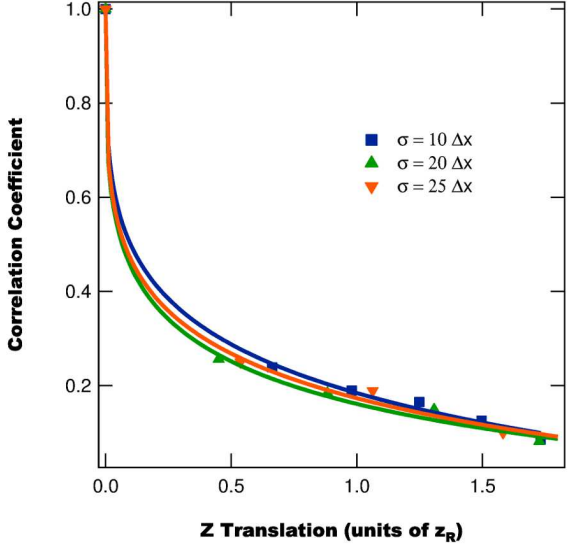}
 \caption{(Color online) Simulated correlation coefficient as a function of sample translation for a bin size of $b=1$ $\Delta$x and three different spot sizes with stretched exponential fits.  The correlation curves are unaffected by the change in spot size.}
 \label{fig:SMZ}
\end{figure}

We also simulate the effect of spot size on the correlation coefficient as a function of $z$ translation, with the results shown in Figure \ref{fig:SMZ}. Note $\sigma=10$ $\Delta$x corresponds to the beam waist.  From Figure \ref{fig:SMZ} the spot size appears to have little, if any, effect on the correlation curves. This result is due to the wavefront-sample coupling depending more on the fractional bin size, rather than the absolute size of the bin on the sample.  In this case, since the bin's fractional size is unchanged by the spot size changing, the change in spot size shouldn't affect the $z$ translational stability.

% This is unexpected as the change in the beam width, $w(z)$, due to a $z$ translation of $\zeta$, at an initial optical axis position $z=z_1$, is given by 
% \begin{align}
%  \frac{dw}{d\zeta}=\frac{w_0^2}{z_R^2}\frac{z_1+\zeta}{w(z_1+\zeta)}, \label{eqn:dwdz}
% \end{align}
% which is found to increase the further you move away from the beam waist and asymptotically approaching
% \begin{align}
%  \frac{dw}{d\zeta}\bigg|_{z_1 \rightarrow \infty}=\frac{w_0}{z_R}. \label{eqn:asymp}
% \end{align}
% Given Equations \ref{eqn:dwdz} and \ref{eqn:asymp} we would therefore expect more $z$ translational stability for the sample being positioned near the beam waist, where $\frac{dw}{dz}$ is smaller, than for the sample positioned further from the beam waist.

\subsection{Rotational Stability}
To simulate the correlation coefficient as a function of rotation we begin first by considering how rotation affects the reflected laser beam.  Due to the nature of the transformations required, in this section we use a continuous coordinate basis to derive the transformations with the understanding that calculations are performed using discrete Fourier Transforms.  In this case the reflection matrix, $\mathbf{r}$, becomes a response function $\mathcal{R}(x-y)$ where the reflected field is given by

\begin{align}
 E'(x)=\int_{-\infty}^\infty dy\mathcal{R}(x-y)E(y)
\end{align}
where $E(y)$ is the incident field. 

\begin{figure}
\centering
\includegraphics{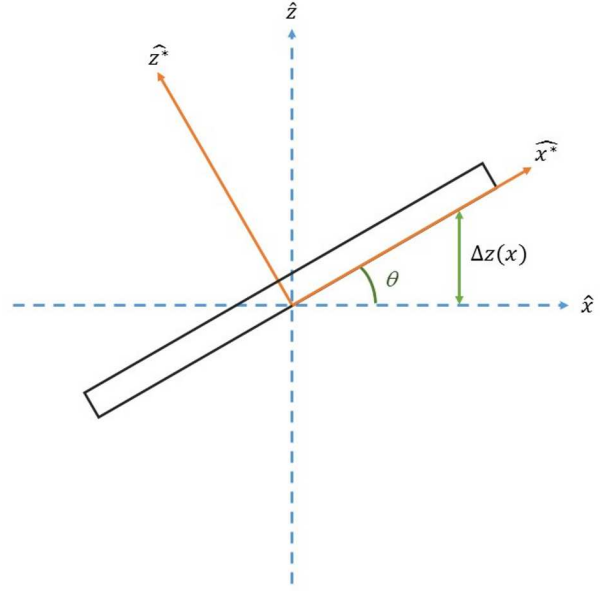}
\caption{(Color online) Coordinates of lab frame and sample frame (denoted by $^*$).}
\label{fig:coor}
\end{figure}

To determine the effects of sample rotation we simulate rotation of the sample around the $y$ axis by an angle $\theta$ as shown in Figure \ref{fig:coor}. Due to the rotation the beam and sample will intersect at different $z$ positions as you move from the center.  Namely, the difference in the intersection position is given by:

\begin{align}
\Delta z(x)=x \tan\theta, \label{eqn:dz}
\end{align}
where $x$ is the transverse position in the lab (unrotated) coordinates. This transverse position dependent offset will result in the addition of a linear phase term to the incident field given by

\begin{align}
\Delta \psi(x)=k\Delta z(x). \label{eqn:dpsi}
\end{align}
The offset also affects the Gaussian beam parameters of the incident field giving a transverse position dependent Gaussian width of

\begin{align}
w(x)=w_0\sqrt{1+\frac{1}{z_R^2}x^2\tan^2\theta},\label{eqn:widthR}
\end{align}

and a transverse position dependent curvature of

\begin{align}
 R(x)=x\tan\theta\left[1+\frac{z_R^2}{x^2\tan^2\theta}\right].\label{eqn:curveR}
\end{align}

Finally, rotating the sample results in coordinate transformation for the sample surface.  The transformation from the sample coordinates (denoted by a superscript *) to the lab coordinates is given by
\begin{align}
x=x^*\cos\theta, \label{eqn:transform}
\end{align}
where $x^*$ is the transverse position on the sample and $x$ is the transverse position in the lab coordinates.  Note that Equation \ref{eqn:transform} implies that for a rotation of $\theta=\pi/2$, all the sample coordinates transform into $x=0$, which is expected as the sample is parallel to the optical axis in this configuration.

With the changes given by Equations \ref{eqn:dpsi}--\ref{eqn:transform} we can write the field reflected from the rotated sample as:

% \begin{widetext}
% \begin{align}
% E'_{r;n^*}=\sum\limits_{m^*=0}^Nr_{n^*m^*}E_0\exp\bigg\{-\frac{(m^*-m_0)^2\cos^2\theta}{\sigma_{m^*\cos\theta+(1+\cos\theta)m_0}^2}-i\psi_{m^*\cos\theta+(1+\cos\theta)m_0}-i\frac{(m^*-m_0)^2\cos^2\theta}{2R_{m^*\cos\theta+(1+\cos\theta)m_0}}\bigg\}
% \end{align}
% \end{widetext}

\begin{widetext}
\begin{align}
E'(x^*)=\int_{-\infty}^{\infty} ds^*\mathcal{R}(x^*-s^*)E_0\exp\bigg\{&-\frac{s^{*2}\cos^2\theta}{w_0^2(1+\frac{1}{z_R^2}s^{*2}\cos^2\theta\tan^2\theta)}-i\psi(s^*\cos\theta) \nonumber
\\&-i\frac{s^{*2}\cos^2\theta}{s^*\cos\theta\tan\theta\left[1+\frac{z_R^2}{s^{*2}\cos^2\theta\tan^2\theta}\right]}-iks^*\sin\theta\bigg\}.\label{eqn:yuck}
\end{align}
 \end{widetext}
 
At this point applying Fraunhoffer diffraction theory will predict the field in a plane parallel to the sample.  However, the detector is setup to be parallel to the unrotated sample. Therefore to account for the coordinate transformation between the sample and lab coordinates we apply angular spectral decomposition to calculate the diffracted field in the detector plane \cite{Matsushima03.01,Ganci81.01,Patorski83.01}. 

Angular spectral decomposition begins by first transforming the electric field reflected from the sample into its spectral representation by taking the Fourier transform of Equation \ref{eqn:yuck}:
\begin{align}
\mathcal{E'}(u^*)=\int_{-\infty}^{\infty}dxE'(x^*)e^{i2\pi u^*x^*}
\end{align}
where $u^*$ is the angular frequency in the sample coordinates. Using a spectral coordinate transform \cite{Matsushima03.01} we can rotate the reflected field into the lab coordinates as

\begin{align}
\mathcal{E'}(v)=\mathcal{E'}(u^*\cos\theta+(\lambda^{-2}-u^{*2})^{1/2}\sin\theta)
\end{align}
where $v$ is the angular frequency in the lab coordinates. Next we calculate the field in the lab's spatial coordinates, which requires taking the Inverse Fourier Transform of the field and the determinate of the coordinate transform's Jacobian, $|J(u^*)|$ \cite{Matsushima03.01}.  The Jacobian term is required to preserve energy conservation as the coordinate transform is nonlinear \cite{Matsushima03.01}. These considerations result in the reflected field in the lab coordinates being,

\begin{align}
E'(x)=\int_{-\infty}^\infty du^* \mathcal{E'}(u^*\cos\theta+[\lambda^{-2}-u^{*2}]^{1/2}\sin\theta) \nonumber
\\ \times |J(u^*)|e^{-i2\pi u^*x},
\end{align}
where the determinate of the Jacobian is given by \cite{Matsushima03.01},

\begin{align}
J(u^*)=\cos\theta-\frac{u^*}{\sqrt{\lambda^{-2}-u^{*2}}}\sin\theta.
\end{align}
With $E'(x)$ calculated we now can calculate the electric field in the detector plane straightforwardly as

\begin{align}
 E_d(x')=\int_{-\infty}^\infty dxE'(x)e^{i\frac{k}{Z}xx'},
\end{align}
where $Z$ is the distance between the sample and detector.

\begin{figure}
 \centering
 \includegraphics{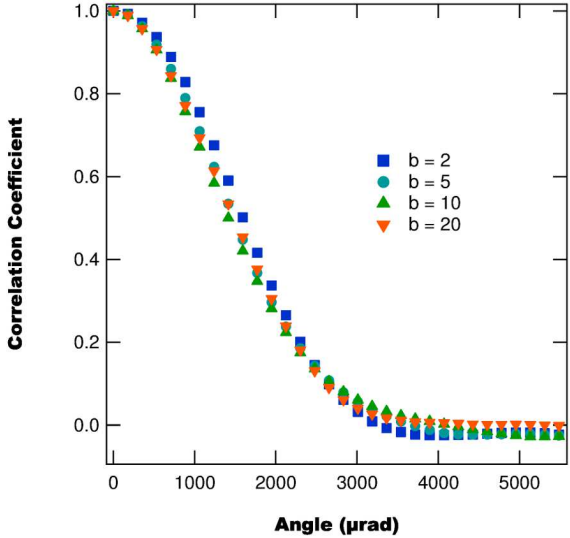}
 \caption{(Color online) Simulated correlation coefficient as a function of sample rotation for a Gaussian width of 200 $\Delta$x and four different bin sizes.  All four curves are found to be within error of each other suggesting that the bin size does not affect rotational stability.}
 \label{fig:MA}
\end{figure}

\begin{figure}
 \centering
 \includegraphics{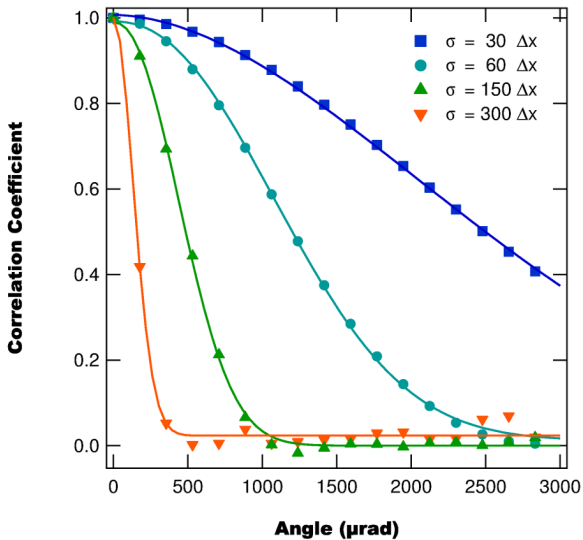}
 \caption{(Color online) Simulated correlation coefficient as a function of sample rotation for a bin size of $b=5$ $\Delta$x and four different spot sizes with stretched exponential fits. The correlation coefficient's HWHM is found to increase with smaller spot size, suggesting that smaller spot sizes are more resistant to sample rotation.}
 \label{fig:SMA}
\end{figure}

Using the model of sample rotation described above -- with the equations converted to a discrete basis for computation -- we calculate the wavefront coupling stability as a function of sample rotation for different bin sizes, shown in Figure \ref{fig:MA}, and different spot sizes, shown in Figure \ref{fig:SMA}.  From Figure \ref{fig:MA} we find that the correlation curve as a function of rotation follows a stretched exponential function with the bin size having no effect on the HWHM. This effect arises due to the relative magnitudes of the changes to the beam due to rotation. For instance, if we assume small angles, such that $\theta^2\approx 0$, Equation \ref{eqn:yuck} becomes,

\begin{align}
E'(x^*)\approx \int_{-\infty}^{\infty} & ds^*\mathcal{R}(x^*-s^*)E_0 \nonumber
\\ &\times \exp\bigg\{-\frac{s^{*2}}{w_0^2}-i\psi(s^*)-iks^*\theta\bigg\},\label{eqn:yuckSim}
\end{align}
where the only remaining influence of rotation is the linear phase term $ks^*\theta$.  As this term only depends on the on sample position, and not the bin size it is easy to see why the bin size does not affect the rotational stability.  To justify this for larger angles, $\theta^2>>0$, we note that at these larger angles the wavefront-sample coupling is already broken by the linear phase term.

% We can understand the lack of bin size dependence by considering how the bin size is transformed due to sample rotation.  Assuming a bin size in lab coordinates of size $b$, the bin in the sample coordinates will be of size,
% 
% \begin{align}
%  b^{*}=\frac{b}{\cos\theta}.
% \end{align}
% Since we consider small rotation angles, such that $\theta^2\approx 0$, the cosine term is essentially unity meaning that $b^{*}=b$. Since the bin size is essentially unchanged during small rotations it will not have an effect on the rotational HWHM. Additionally, for larger rotation angles -- where the bin size is different in lab and sample coordinates -- the wavefront-sample coupling is already broken.

While the coupling's rotational stability is found to be independent of bin size, from Figure \ref{fig:SMA} we find that the spot size has a drastic effect on the coupling's stability, namely, that as the spot size increases the coupling stability decreases. This effect is a direct result of Equation \ref{eqn:dz}, which shows that the scale of the $z$ offset is linearly related to the transverse position, meaning that for larger spot sizes there will be larger $z$ offsets, resulting in greater phase changes. 

\section{Method}
To experimentally determine the coupling stability of an optimized wavefront under translation and rotation we use an optimized reflection setup and ZrO$_2$ doped polyurethane (PU) nanocomposites. The preparation of the ZrO$_2$/PU nanocomposites is as follows. Spherical zirconia nanoparticles, ZrO$_2$ NPs, are synthesized by forced hydrolysis followed by calcination at temperature and time of 600$^{\circ}$C and 1 h, respectively \cite{Gunawidjaja13.01}.   The ZrO$_2$ NPs are then hydrophobized by boiling the dispersion of the ZrO$_2$ NPs (2 mg/mL) in 1 vol\% solution of n-octadecyltriethoxysilane (Gelest, SIO6642.0) in toluene \cite{Gunawidjaja11.01}. The hydrophobized ZrO$_2$ NPs are dispersed in 10 wt\% bisphenol A diglycidyl ether, BADGE (EPONTM resin 828, Miller-Stephenson Chemical Company, Inc.) solution in toluene at a 20 mg/mL concentration with the aid of sonication. Toluene acts as a thinner for the otherwise viscous BADGE fluid. The remaining BADGE is then added to yield the desired weight fraction 
of ZrO$_2$ NPs. The mixture is further sonicated until a homogeneous mixture is achieved and the toluene is evaporated in vacuum.  Once the BADGE mixture is prepared an equivalent amount of diethylene triamine, DETA (Epikure 3223 curing agent, Miller-Stephenson Chemical Company, Inc.) is added and the mixture is mixed thoroughly and then poured onto a rectangular glass slide (1''x1.5''). The polymer is then polymerized in an 80 $^{\circ}$C oven for 2 hours with the resulting films having thickness of 1-3 mm and a scattering length of approximately 4 $\mu$m.  This implies that multiple scattering events occur within the sample when illuminated.

\begin{figure}
 \centering
 \includegraphics{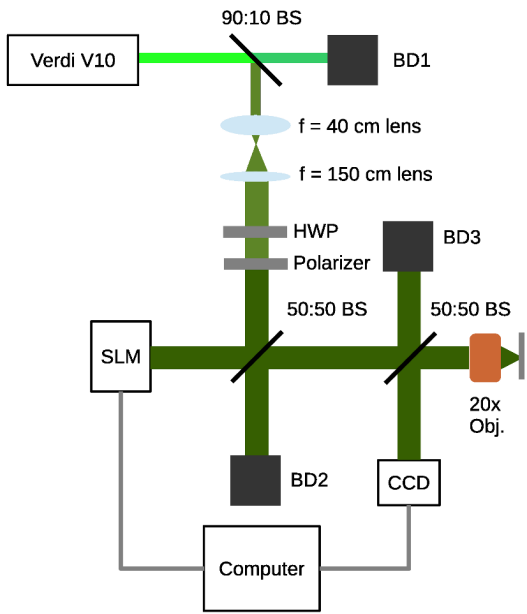}
 \caption{(Color online) Schematic of optimized reflection setup.}
 \label{fig:setup}
\end{figure}

With the samples prepared we place them inside of an optimized reflection setup consisting of: a Coherent Verdi Nd:YVO$_4$ CW laser, a Boulder Nonlinear Sciences LCOS-SLM, a Thorlabs CMOS camera and various optics.  See Figure \ref{fig:setup} for a schematic of the experimental setup.  The Verdi is operated at full power (10 W), where the laser is most stable.  Since the 10 W output is well above the damage threshold of the SLM we use a 90:10 beamsplitter to pick off 1 W of the laser, which is then expanded and passed through a half-waveplate (HWP) polarizer pair to further control the beam intensity, as well as maintain the correct polarization for the SLM.  The beam is then reflected off the SLM and focused onto the sample using a 20$\times$ high working distance objective. The resulting speckle pattern reflected from the sample is then collected by the same objective and reflected onto the Thorlabs camera.  The camera and SLM are interfaced to a computer which uses a microgenetic algorithm \cite{
Anderson15.02} to optimize the wavefront such that the reflected light produces a spot focus on the camera array.

To translate the samples accurately we use a Thorlabs flexure stage with differential micrometers giving a translational sensitivity of 500 nm.  In order to minimize the effect of translational cross talk we set the sample zero position to be near the middle of the full micrometer range, as per the manufacturers instructions.  For rotating the sample we use a Newport precision rotation stage with a sensitivity of 30 arc seconds (145.4 $\mu$rad)

The procedure for measuring the positional stability of the wavefront coupling is as follows.  First, the optimal wavefront, which produces the largest intensity in the target area, is determined using a microgenetic optimization algorithm \cite{Anderson15.02}.  Once the optimal wavefront is found the target intensity profile is imaged four times and averaged to find the optimized intensity profile.  The sample is then translated or rotated a fixed distance and the new intensity profile is measured four times and averaged to find the intensity profile corresponding to the new position.  This process is then repeated until the intensity in the target area is the same as the background, which corresponds to full decoupling of the wavefront.  Once the intensity profiles are measured they are each compared to the optimized intensity profile using Pearson's correlation coefficient, Equation \ref{eqn:corr}.

\section{Results and Discussion}  

\subsection{Bin Size}
The first experimental parameter we vary when measuring the translational and rotational wavefront coupling stability is the SLM bin size. For these measurements we use a sample with a NP concentration of 25 wt\%, three different bin sizes, $b=\{8,12,18\}$, and a spot size of $2w_0=2.2$ $\mu$m.

\begin{figure}
 \centering
 \includegraphics{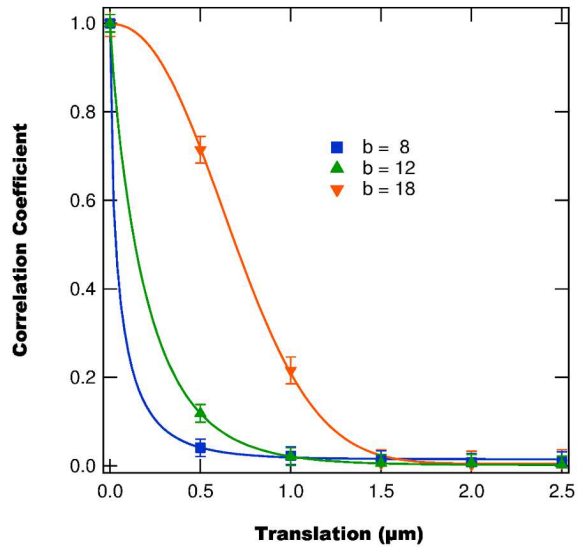}
 \caption{(Color online) Measured correlation coefficient as a function of sample translation in the transverse direction for a 25 wt\% sample, spot size of 2.2 $\mu$m and three different bin sizes with stretched exponential fits.  Like the simulated results the fit width is found to decrease as the bin size decreases. }
 \label{fig:X}
\end{figure}

We begin by measuring the correlation coefficient as a function transverse translation for the three different bin sizes as shown in Figure \ref{fig:X}, with the data fit to Equation \ref{eqn:strexp}. From Figure \ref{fig:X} we see that as the bin size increases the HWHM of the correlation curves increases, as tabulated in Table \ref{tab:bin}. Additionally from Figure \ref{fig:X} we see that the correlation curves change shape as the bin size increases with the small bin sizes having much steeper decays, corresponding to $\beta<1$, and the larger bin sizes having more gradual decays, corresponding to $\beta>1$.  Both these results, the change in HWHM and $\beta$ with bin size, are consistent with simulation.

\begin{figure}
 \centering
 \includegraphics{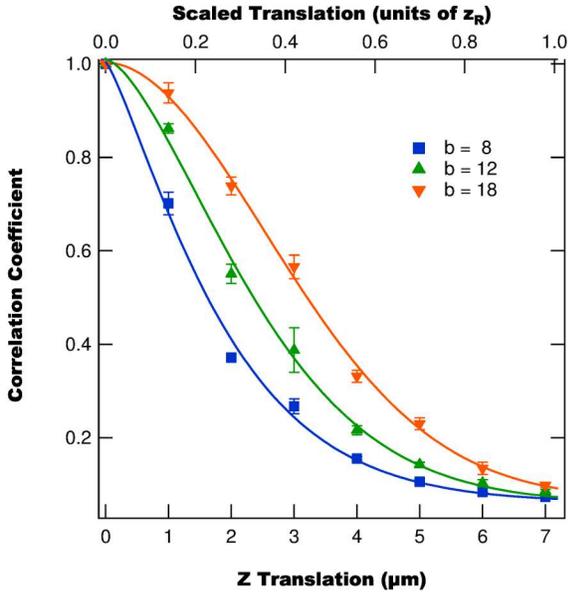}
 \caption{(Color online) Measured correlation coefficient as a function of sample translation along the optical axis for a 25 wt\% sample, spot size of 2.2 $\mu$m and three different bin sizes with fits to stretched exponentials.}
 \label{fig:Z}
\end{figure}

After translating the sample in the transverse direction we next translate the sample along the optical axis.  Figure \ref{fig:Z} shows the correlation coefficient as a function of $z$ position for the different bin sizes tested.  We find that as the bin size increases the coupling stability increases, with the HWHM becoming larger as listed in Table \ref{tab:bin}.  The curves are also found to intersect near zero as the z translation approaches a value equal to the Rayleigh range.  These results, the HWHM increase and curve intersection, are predicted by simulations. However, the intersection point in the simulations is closer to $z=1.8$ $z_R$, while the experimental intersection is closer to $z=z_R$.

At this point it is important to note the relative stability lengths of transverse and optical axis translation.  From Table \ref{tab:bin} we find that for the smallest bin size the transverse HWHM is only 37.3 nm while the $z$ HWHM is 1364 nm, corresponding to a scale difference of 36.6 $\times$.  This scale difference, also observed in the simulations, implies that the wavefront-sample coupling is more sensitive to transverse translation than to optical axis translation. The underlying mechanism of this difference is related to the different effects each translation produces.  Transverse translation results in the center of the beam changing position on the sample surface, thereby effectively shifting the fields index, but leaving the reflection matrix's indices the same.  This shift quickly breaks the coupling between the optimal wavefront and the sample.  On the other hand, for translation along the optical axis the center of the beam is unchanged and only the beam width shifts.  This means that the 
field's positional indices and the reflection matrix's indices match for a longer translation, resulting in the increased positional stability.

\begin{figure}
 \centering
 \includegraphics{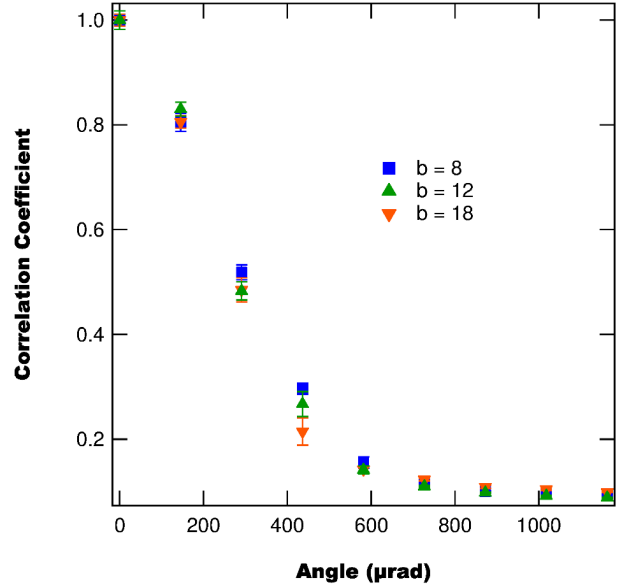}
 \caption{(Color online) Measured correlation coefficient as a function of sample rotation for a 25 wt\% sample, spot size of 2.2 $\mu$m and three different bin sizes.  The correlation curves for each bin size are found to be within experimental uncertainty of each other, suggesting that bin size does not play a role in the rotational stability.}
 \label{fig:A}
\end{figure}

Next we measure the correlation coefficient during sample rotation.  Figure \ref{fig:A} shows the correlation coefficient as a function of rotation angle for the three different bin sizes, with all three correlation curves found to be within uncertainty of each other.  Fitting the curves to Equation \ref{eqn:strexp} we determine their HWHMs, reported in Table \ref{tab:bin}, and find that the HWHMs are unchanged as a function of bin size.  These results are consistent with simulation where the correlation coefficient as a function of angle follows a Gaussian with the width being unaffected by bin size.

\begin{table*}
 \centering
 \begin{tabular}{|c|c|c|c|}
  \hline
  \textbf{Bin Size} & \textbf{X HWHM (nm)} & \textbf{Z HWHM (nm)} & \textbf{Rot. HWHM ($\mu$rad)} \\
  \hline
  18  & 693.6 $\pm$ 9.0 & 2844 $\pm$ 26  & 254 $\pm$ 12 \\
  12 & 137.8 $\pm$ 5.9 & 2228 $\pm$ 32  & 261 $\pm$ 12 \\
  8 & 37.3  $\pm$ 3.9 & 1364 $\pm$ 22  & 265.1 $\pm$ 6.2 \\
  \hline
 \end{tabular}
\caption{Translational and rotational HWHMs for different bin sizes. For both $x$ and $z$ translation smaller bin sizes yield larger HWHM, while the rotation HWHM is unchanged, within experimental uncertainty, for each bin size.}
\label{tab:bin}
\end{table*}

\subsection{Spot Size}
The next experimental parameter (that can affect the correlation as a function of translation and rotation) which we vary is the beam spot size. These measurements are performed using a 25 wt\% NP concentration sample, a bin size of $b=16$, and four different spot sizes having Gaussian widths of $\sigma=$ 1.1 $\mu$m, 60 $\mu$m, 240 $\mu$m, and 540 $\mu$m.  Note that the beam waist for the system corresponds to $\sigma=1.1$ $\mu$m, or a spot size of $2w_0=2.2$ $\mu$m.

\begin{figure}
 \centering
 \includegraphics{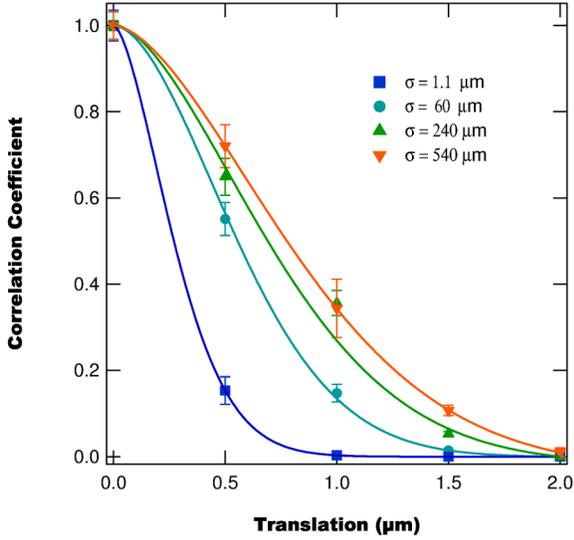}
 \caption{(Color online) Measured correlation coefficient as a function of sample translation in the transverse direction for a 25 wt\% sample, bin size of $b=16$, and four different spot sizes.}
 \label{fig:SX}
\end{figure}

For each spot size we first translate the sample in the transverse direction and measure the correlation coefficient at each position, as shown in Figure \ref{fig:SX}.  From Figure \ref{fig:SX} we find that as the spot size increases the correlation curves broaden (the HWHM are listed in Table \ref{tab:spot}), which is consistent with simulation.  The underlying mechanism of this increase in stability is due to the larger spot size causing transverse translations to have a smaller effect on the incident wavefront, as the fractional change in wavefront with transverse translation is inversely proportional to the spot size.

\begin{figure}
 \centering
 \includegraphics{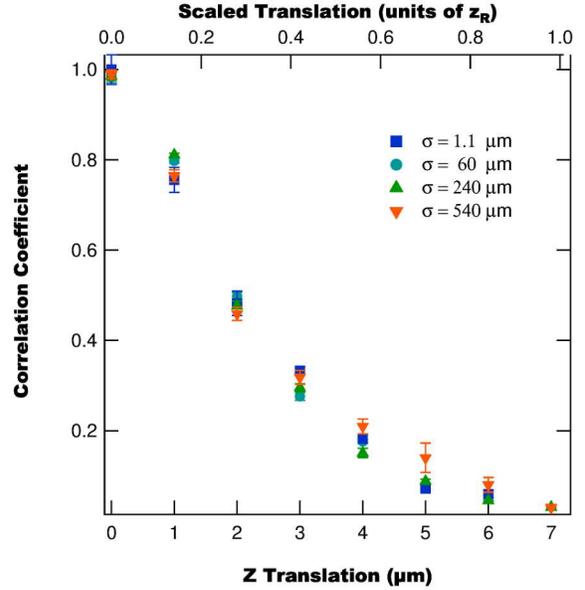}
 \caption{(Color online) Measured correlation coefficient as a function of sample translation along the optical axis for a 25 wt\% sample, bin size of $b=16$, and four different spot sizes.}
 \label{fig:SZ}
\end{figure}

We next measure the effect of changing the spot size on the correlation coefficient as a function of optical axis position.  Figure \ref{fig:SZ} shows the correlation coefficient as a function of $z$ translation for the four different spot sizes, with all four curves overlapping within experimental uncertainty.  This overlap results in the HWHM of each curve, see Table \ref{tab:spot}, being within uncertainty of each other, which is consistent with simulation. As discussed in the simulation section, this effect is due to the bin size to spot size ratio being unchanged with changing spot size.  This implies that the more important quantity to the translational stability is the fractional bin size, and not the absolute bin size.

\begin{figure}
 \centering
 \includegraphics{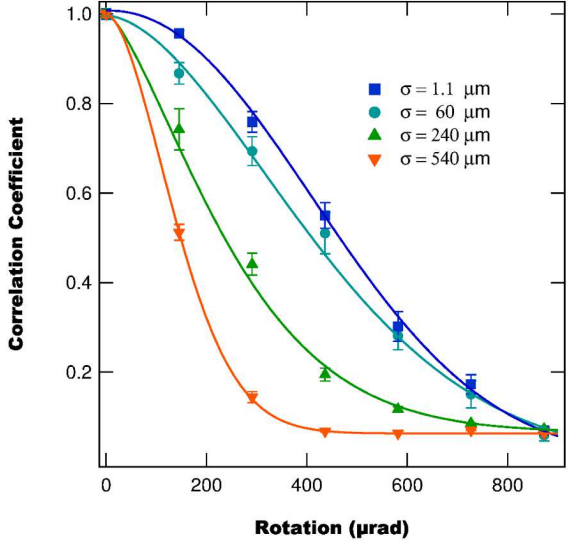}
 \caption{(Color online) Measured correlation coefficient as a function of sample rotation for a 25 wt\% sample, bin size of $b=16$, and four different spot sizes.}
 \label{fig:SA}
\end{figure}

After measuring the translational coupling stability with different spot sizes we move on to measuring the effect of the spot size on the correlation coefficient as a function of rotation as shown in Figure \ref{fig:SA}. From Figure \ref{fig:SA} we find that as the spot size increases the correlation curves narrow, see Table \ref{tab:spot} for the HWHMs, and the initial decay steepens, with the stretch parameter approaching $\beta=1$. These results are consistent with simulation, with the underlying mechanism once again related to larger spot sizes producing larger wavefront distortions than small spot sizes.

\begin{table*}
 \centering
 \begin{tabular}{|c|c|c|c|}
  \hline
  \textbf{Spot Size ($\mu$m)} & \textbf{X HWHM (nm)} & \textbf{Z HWHM (nm)} & \textbf{Rot. HWHM ($\mu$rad)} \\
  \hline
  1.1  & 270.6 $\pm$ 9.4 & 2045 $\pm$ 40  & 462   $\pm$ 10   \\
  60   & 557   $\pm$ 23  & 2048 $\pm$ 50  & 411   $\pm$ 12   \\
  240  & 699   $\pm$ 81  & 2040 $\pm$ 25  & 219.8 $\pm$ 7.3  \\
  540  & 796  $\pm$ 28  & 2050 $\pm$ 44  & 140.9 $\pm$ 3.5  \\
  \hline
 \end{tabular}
\caption{Translational and rotational HWHMs for different incident spot sizes.}
\label{tab:spot}
\end{table*}

\subsection{Nanoparticle Concentration}
Thus far we have considered the effects of bin size and spot size on the wavefront-sample coupling's ability to withstand sample rotation and translation, with the results found to be consistent with simulations using the extended RPGBM. The final experimental parameter we consider is the NP concentration of the samples, which is not modeled by the extended RPGBM.  Therefore we do not have any simulation results to make comparisons too.  In order to effectively model the effect of NP concentration the transmission and reflection matrices need to be calculated in such a way as to take scatterer concentration into account, which cannot be done by the random unitary matrix model used by the extended RPGBM.  Work is currently underway to address this limitation, but is beyond the scope of this study.

\begin{figure}
 \centering
 \includegraphics{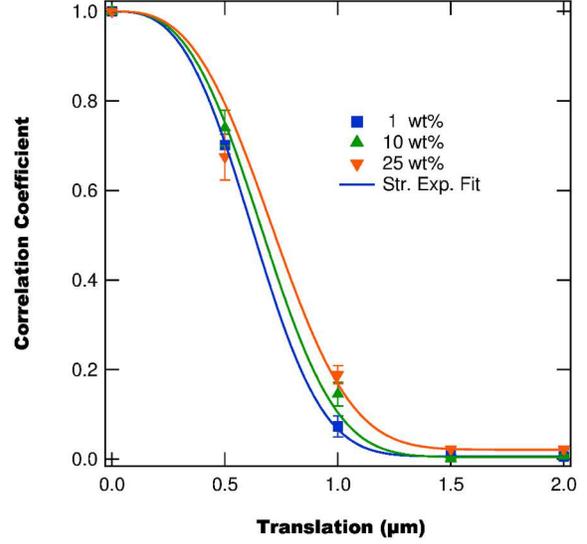}
 \caption{(Color online) Measured correlation coefficient as a function of sample translation in the transverse direction for a spot size of 2.2 $\mu$m, a bin size of $b=20$, and three different NP concentrations with stretched exponential fits.  As the concentration increases the width increases.}
 \label{fig:CX}
\end{figure}

To measure the effect of NP concentration on the wavefront-sample coupling's rotational and translational stability we use three different samples with concentrations of 1 wt\%, 10 wt\%, and 25 wt\%, a bin size of $b=20$, and a spot size of $2\sigma=2.2$ $\mu$m.  We first measure the effect of the different NP concentrations on the coupling's transverse translation stability with the correlation coefficient data and stretched exponential fits shown in Figure \ref{fig:CX}.  From Figure \ref{fig:CX} we find that the correlation curve width increases with increasing concentration, albeit only to a small degree. Table \ref{tab:conc} lists the HWHMs for the three concentrations, from which we see that the increase in HWHM is larger than experimental uncertainty, but is relatively small compared to the changes due to the bin size and spot size. The increase in the translational stability with increasing concentration is due to the high concentration samples having more uniform surfaces, such that there is less 
variation in the reflection matrix as the sample is translated in the transverse direction.

\begin{figure}
 \centering
 \includegraphics{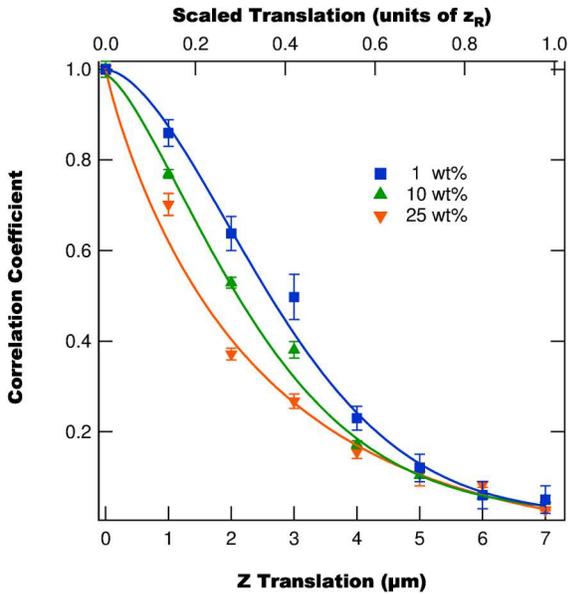}
 \caption{(Color online) Measured correlation coefficient as a function of sample translation along the optical axis for a spot size of 2.2 $\mu$m, a bin size of $b=20$, and three different concentrations with stretched exponential fits.}
 \label{fig:CZ}
\end{figure}

While the effect of NP concentration is found to be small on the wavefront-sample coupling's transverse translational stability, its influence is more apparent when we consider translation along the optical axis.  Figure \ref{fig:CZ} shows the correlation coefficent as a function of $z$ position and stretched exponential fits for the three concentrations.  From Figure \ref{fig:CZ} we find that the three curves intersect near the Rayleigh range (also observed for varying the bin size and spot size) and that as the concentration increases the HWHM (listed in Table \ref{tab:conc}) of the curves decreases, implying that the lower concentration samples are more stable to translation along the optical axis. 

The mechanism behing the increased stability of low concentration samples to translation along the optical axis is related to the relative light penetration depths of the different samples.  As the concentration increases, the scattering length decreases meaning that light penetrates less deeply into the sample.  This depth can be viewed as defining an effective interaction volume.  For lower concentration samples this volume is larger than that of higher concentrations.  Therefore when translating the sample along the optical axis, the fractional change in the scattering volume will be smaller for large scattering volumes than for small scattering volumes. This difference implies that the wavefront-sample interaction will be more affected in the higher concentration samples, than in the low concentration samples, which is observed.

\begin{figure}[b]
 \centering
 \includegraphics{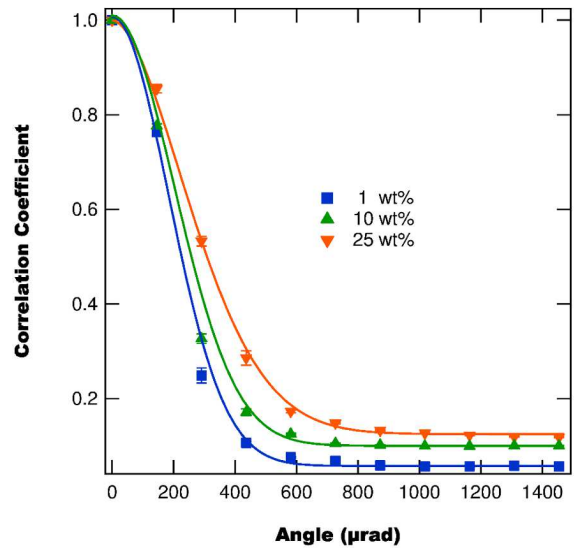}
 \caption{(Color online) Measured correlation coefficient as a function of sample rotation for a spot size of 2.2 $\mu$m, a bin size of $b=20$, and three different sample concentrations with Gaussian fits.  The half-width angle decreases as a function of increasing concentration, suggesting that higher concentrations are more stable under rotation.}
 \label{fig:CA}
\end{figure}

Lastly, we measure the effect of the NP concentration on the wavefront-sample coupling's rotational stability.  Figure \ref{fig:CA} shows the correlation coefficient as a function of sample rotation for the three different concentrations and fits to stretched exponentials.  Similar to the transverse translation we find that higher concentrations result in increased rotational stability, with the HWHM, tabulated in Table \ref{tab:conc} increasing as a function of concentration. The effect of increasing concentration is also found to be small, with the HWHM only varying by $\approx 60$ $\mu$rad between the 1 wt\% and 25 wt\% sample. This difference is due to the surface being more uniform for the high concentration samples than for the low concentration samples.

\begin{table*}[t]
 \centering
 \begin{tabular}{|c|c|c|c|}
  \hline
  \textbf{Concentration} & \textbf{X Half-Width (nm)} & \textbf{Z Half-Width (nm)} & \textbf{Rot. Half-Width ($\mu$rad)} \\
  \hline
  1  &  628 $\pm$ 17 & 2582 $\pm$ 80 & 215.6 $\pm$ 1.6 \\
  10 &  670 $\pm$ 22 & 2090 $\pm$ 51 & 237.1 $\pm$ 1.4 \\
  25 &  715 $\pm$ 16 & 1365 $\pm$ 25 & 276.4 $\pm$ 2.7 \\
  \hline
 \end{tabular}
\caption{Translational and rotational HWHMs for different NP concentrations}
\label{tab:conc}
\end{table*}

Apart from the correlation curve's HWHM we also note from Figure \ref{fig:CA} that the correlation offset ($c_0$ in Equation \ref{eqn:corr}) is found to increase with increasing concentration.  This result is due to the nature of the speckle pattern as a function of concentration.  As the concentration increases the speckle grain is found to decrease (i.e. individual speckles become smaller) resulting in a larger speckle density. If we compare two different regions of the speckle pattern we find that the similarity of the two regions increases with the speckle density.  This implies that comparing two images with high speckle density will result in more similarity and therefore a larger background correlation coefficient.

\section{Conclusions}
In the above sections we have explored theoretically and experimentally the effects of three different system parameters (bin size, spot size, and NP concentration) on the wavefront-sample coupling's translational and rotational stability for an optimized reflection system.  We find that the simulations -- using a model developed from the random phase Gaussian beam model \cite{Anderson14.06} and Vellekoop's random unitary matrix model \cite{Vellekoop08.01} -- are within qualitative agreement with experimental results. The results of experiment and simulation are summarized below.

When changing bin sizes it is found that larger bins provide more translational stability in both the transverse and $z$ directions.  This improvement is due to the fractional translation change being inversely proportional to the bin size, which means that larger bins result in the on sample wavefront changing less during translation than for smaller bins.  While the bin size affects the translational stability it does not affect the wavefront-sample coupling's rotational stability. This is expected as: 1) the equations describing the effect of sample rotation (Equations \ref{eqn:dpsi}--\ref{eqn:transform}) are independent of the bin size and 2) for small angles the bin size is invariant under rotation.

The next system parameter varied, the on sample spot size, is found to influence both the transverse translation and rotational stability of the wavefront-sample coupling, but not the stability of translation along the optical axis.  As the spot size is increased the transverse translational stability is found to increase, while the rotational stability decreases.  The benefit of larger spot sizes on the transverse stability is due to the fractional wavefront change (during transverse translation) being inversely proportional to the spot size.  As the spot size increases the effect of translating in the transverse direction has a smaller effect on the wavefront seen by the sample.  However, in the case of rotation, the increased spot size results in larger wavefront changes due to Equation \ref{eqn:yuck}'s dependence on the spot size. Finally, the lack of an effect on translation along the optical axis from changing the spot size is due to the $z$ translational stability depending on the 
fractional bin size, and not the absolute bin size. While the absolute bin size changes with spot size, the fractional bin size is unchanged, thus explaining the observed effect.

The last system parameter varied, the sample's NP concentration, is found to increase the wavefront-sample's coupling stability to transverse translation and rotation when increased, while also decreasing the stability to translations along the optical axis. These results are due to higher concentration samples having more surface uniformity and smaller scattering volumes.

The sum total of these results suggest that in practice a balance needs to be struck in order to obtain the best possible positional stability. This observation is due to both spot size and sample NP concentration being found to produce conflicting results (e.g. smaller spot sizes increase rotational stability while decreasing transverse stability).  Additionally, changing these parameters for stability needs to be weighed against their effects on the maximum possible enhancement, described in Ref. \cite{Anderson14.06}.  For instance, larger bin sizes are more positionally stable but produce smaller enhancements. Therefore in some applications it may be more important to sacrifice stability for larger enhancements.

\section{Acknowledgements}
This work was supported by the Defense Threat Reduction Agency, Award \# HDTRA1-13-1-0050 to Washington State University.

\bibliographystyle{osajnl}
\bibliography{PrimaryDatabase,ASLbib}

\end{document}